\documentclass[twocolumn]{aastex63}
\received{2020, March 7}
\revised{2020, April 20}
\accepted{2020, April 20}
\published{2020 May 7}
\usepackage{graphicx,amsmath,amssymb,amsfonts, soul,color}
\usepackage{hyperref}
\newcommand{\NWU}{Centre for Space Research, North-West University, Potchefstroom 2520, South Africa}
\newcommand{\HH}{Universit\"at Hamburg, Institut f\"ur Experimentalphysik, Luruper Chaussee 149, D 22761 Hamburg, Germany}
\newcommand{\MPIK}{Max-Planck-Institut f\"ur Kernphysik, P.O. Box 103980, D 69029 Heidelberg, Germany}
\newcommand{\DIAS}{Dublin Institute for Advanced Studies, 31 Fitzwilliam Place, Dublin 2, Ireland}
\newcommand{\RAU}{High Energy Astrophysics Laboratory, RAU,  123 Hovsep Emin St  Yerevan 0051, Armenia}
\newcommand{\YPI}{Yerevan Physics Institute, 2 Alikhanian Brothers St., 375036 Yerevan, Armenia}
\newcommand{\HUB}{Institut f\"ur Physik, Humboldt-Universit\"at zu Berlin, Newtonstr. 15, D 12489 Berlin, Germany}
\newcommand{\UNAM}{University of Namibia, Department of Physics, Private Bag 13301, Windhoek, Namibia, 12010}
\newcommand{\GRAPPA}{GRAPPA, Anton Pannekoek Institute for Astronomy, University of Amsterdam,  Science Park 904, 1098 XH Amsterdam, The Netherlands}
\newcommand{\Linnaeus}{Department of Physics and Electrical Engineering, Linnaeus University,  351 95 V\"axj\"o, Sweden}
\newcommand{\RUB}{Institut f\"ur Theoretische Physik, Lehrstuhl IV: Weltraum und Astrophysik, Ruhr-Universit\"at Bochum, D 44780 Bochum, Germany}
\newcommand{\LFUI}{Institut f\"ur Astro- und Teilchenphysik, Leopold-Franzens-Universit\"at Innsbruck, A-6020 Innsbruck, Austria}
  \newcommand{\Adelaide}{School of Physical Sciences, University of Adelaide, Adelaide 5005, Australia}
\newcommand{\LUTH}{LUTH, Observatoire de Paris, PSL Research University, CNRS, Universit\'e Paris Diderot, 5 Place Jules Janssen, 92190 Meudon, France}
\newcommand{\LPNHE}{Sorbonne Universit\'e, Universit\'e Paris Diderot, Sorbonne Paris Cit\'e, CNRS/IN2P3, Laboratoire de Physique Nucl\'eaire et de Hautes Energies, LPNHE, 4 Place Jussieu, F-75252 Paris, France}
\newcommand{\LUPM}{Laboratoire Univers et Particules de Montpellier, Universit\'e Montpellier, CNRS/IN2P3,  CC 72, Place Eug\`ene Bataillon, F-34095 Montpellier Cedex 5, France}
\newcommand{\IRFU}{IRFU, CEA, Universit\'e Paris-Saclay, F-91191 Gif-sur-Yvette, France}
\newcommand{\UWarsaw}{Astronomical Observatory, The University of Warsaw, Al. Ujazdowskie 4, 00-478 Warsaw, Poland}
\newcommand{\CPPM}{Aix Marseille Universit\'e, CNRS/IN2P3, CPPM, Marseille, France}
\newcommand{\IFJPAN}{Instytut Fizyki J\c{a}drowej PAN, ul. Radzikowskiego 152, 31-342 Krak{\'o}w, Poland}
\newcommand{\WITS}{School of Physics, University of the Witwatersrand, 1 Jan Smuts Avenue, Braamfontein, Johannesburg, 2050 South Africa}
\newcommand{\LAPP}{Laboratoire d'Annecy de Physique des Particules, Univ. Grenoble Alpes, Univ. Savoie Mont Blanc, CNRS, LAPP, 74000 Annecy, France}
\newcommand{\LSW}{Landessternwarte, Universit\"at Heidelberg, K\"onigstuhl, D 69117 Heidelberg, Germany}
\newcommand{\CENBG}{Universit\'e Bordeaux, CNRS/IN2P3, Centre d'\'Etudes Nucl\'eaires de Bordeaux Gradignan, 33175 Gradignan, France}
\newcommand{\IAAT}{Institut f\"ur Astronomie und Astrophysik, Universit\"at T\"ubingen, Sand 1, D 72076 T\"ubingen, Germany}
\newcommand{\LLR}{Laboratoire Leprince-Ringuet, École Polytechnique, CNRS, Institut Polytechnique de Paris, F-91128 Palaiseau, France}
\newcommand{\APC}{APC, AstroParticule et Cosmologie, Universit\'{e} Paris Diderot, CNRS/IN2P3, CEA/Irfu, Observatoire de Paris, Sorbonne Paris Cit\'{e}, 10, rue Alice Domon et L\'{e}onie Duquet, 75205 Paris Cedex 13, France}
\newcommand{\Leicester}{Department of Physics and Astronomy, The University of Leicester, University Road, Leicester, LE1 7RH, United Kingdom}
\newcommand{\NCAC}{Nicolaus Copernicus Astronomical Center, Polish Academy of Sciences, ul. Bartycka 18, 00-716 Warsaw, Poland}
\newcommand{\UP}{Institut f\"ur Physik und Astronomie, Universit\"at Potsdam,  Karl-Liebknecht-Strasse 24/25, D 14476 Potsdam, Germany}
\newcommand{\ECAP}{Friedrich-Alexander-Universit\"at Erlangen-N\"urnberg, Erlangen Centre for Astroparticle Physics, Erwin-Rommel-Str. 1, D 91058 Erlangen, Germany}
\newcommand{\DESY}{DESY, D-15738 Zeuthen, Germany}
\newcommand{\UJK}{Obserwatorium Astronomiczne, Uniwersytet Jagiello{\'n}ski, ul. Orla 171, 30-244 Krak{\'o}w, Poland}
\newcommand{\NCUT}{Centre for Astronomy, Faculty of Physics, Astronomy and Informatics, Nicolaus Copernicus University,  Grudziadzka 5, 87-100 Torun, Poland}
\newcommand{\UFS}{Department of Physics, University of the Free State,  PO Box 339, Bloemfontein 9300, South Africa}
\newcommand{\Rikkyo}{Department of Physics, Rikkyo University, 3-34-1 Nishi-Ikebukuro, Toshima-ku, Tokyo 171-8501, Japan}
\newcommand{\KAVLI}{Kavli Institute for the Physics and Mathematics of the Universe (WPI), The University of Tokyo Institutes for Advanced Study (UTIAS), The University of Tokyo, 5-1-5 Kashiwa-no-Ha, Kashiwa, Chiba, 277-8583, Japan}
\newcommand{\Tokyo}{Department of Physics, The University of Tokyo, 7-3-1 Hongo, Bunkyo-ku, Tokyo 113-0033, Japan}
\newcommand{\RIKKEN}{RIKEN, 2-1 Hirosawa, Wako, Saitama 351-0198, Japan}
\newcommand{\Oxford}{University of Oxford, Department of Physics, Denys Wilkinson Building, Keble Road, Oxford OX1 3RH, UK}
\newcommand{\CerrutiNowAt}{Now at Institut de Ci\`{e}ncies del Cosmos (ICC UB), Universitat de Barcelona (IEEC-UB), Mart\'{i} Franqu\`es 1, E08028 Barcelona, Spain}

\newcommand{\RNum}[1]{\uppercase\expandafter{\romannumeral #1\relax}}

\begin{document}

\title{Probing the magnetic field in the GW170817 outflow using H.E.S.S. observations}

\correspondingauthor{H.~Ashkar, S.~Ohm, Q.~Piel, F.~Sch\"ussler, A.M.~Taylor, X.~Rodrigues}
\email{contact.hess@hess-experiment.eu}

% Author list ApJ style
\author{H.~Abdalla} \affiliation{\NWU}
\author{R.~Adam} \affiliation{\LLR}
\author{F.~Aharonian} \affiliation{\MPIK}\affiliation{\DIAS}\affiliation{\RAU}
\author{F.~Ait~Benkhali} \affiliation{\MPIK}
\author{E.O.~Ang\"uner} \affiliation{\CPPM}
\author{M.~Arakawa} \affiliation{\Rikkyo}
\author{C.~Arcaro} \affiliation{\NWU}
\author{C.~Armand} \affiliation{\LAPP}
\author{T.~Armstrong} \affiliation{\Oxford}
\author{H.~Ashkar} \affiliation{\IRFU}
% ...
\collaboration{275}{(H.E.S.S. Collaboration)}
\author{X.~Rodrigues} \affiliation{DESY, D-15738 Zeuthen, Germany}
\nocollaboration{1}

\begin{abstract}
The detection of the first electromagnetic counterpart to the binary neutron star (BNS) merger remnant GW170817 established the connection between short $\gamma$-ray bursts and BNS mergers. It also confirmed the forging of heavy elements in the ejecta (a so-called {\it kilonova}) via the {\it r-process nucleosynthesis}. The appearance of non-thermal radio and X-ray emission, as well as the brightening, which lasted more than 100 days, were somewhat unexpected. Current theoretical models attempt to explain this temporal behavior as either originating from a relativistic off-axis jet or a kilonova-like outflow. In either scenario, there is some ambiguity regarding how much energy is transported in the non-thermal electrons versus the magnetic field of the emission region. Combining the VLA (radio) and Chandra (X-ray) measurements with observations in the GeV-TeV domain can help break this ambiguity, almost independently of the assumed origin of the emission. Here we report for the first time on deep H.E.S.S. observations of GW170817 / GRB~170817A between 124 and 272 days after the BNS merger with the full H.E.S.S. array of telescopes, as well as on an updated analysis of the prompt ($<$5 days) observations with the upgraded H.E.S.S. phase-I telescopes. We discuss implications of the H.E.S.S. measurement for the magnetic field in the context of different source scenarios.
 \end{abstract}


%\maketitle 

\section{Introduction}
\label{sec:introduction}  
The Gravitational Wave (GW) event detected on August
17$^{\mathrm{th}}$ 2017 by the advanced LIGO and Virgo detectors
resulted from the merger of two neutron stars (NS). The GW signal was
followed after $\sim$2 seconds by a short, low-luminosity $\gamma$-ray
burst (GRB) and seen by the Fermi-GBM \citep{GBM2017} and INTEGRAL-SPI
\citep{Integral2017} instruments. Observations in the optical band
later associated this GRB (GRB~170817A) as the counterpart of GW170817
and localized it to the host galaxy NGC\,4993 \citep{Coulter2017a}.
The fading UV, optical, and infrared radiation was followed by a
rising non-thermal radio and X-ray signal after $\sim$9 days
\citep{Troja2017}. This behavior, as predicted by \citet{Takami2014},
is indicative of efficient particle acceleration in the NS merger
remnant and subsequent synchrotron emission of accelerated electrons
in the ejecta's magnetic field.

\begin{figure}[!h]
\centering
\includegraphics[width=0.5\textwidth]{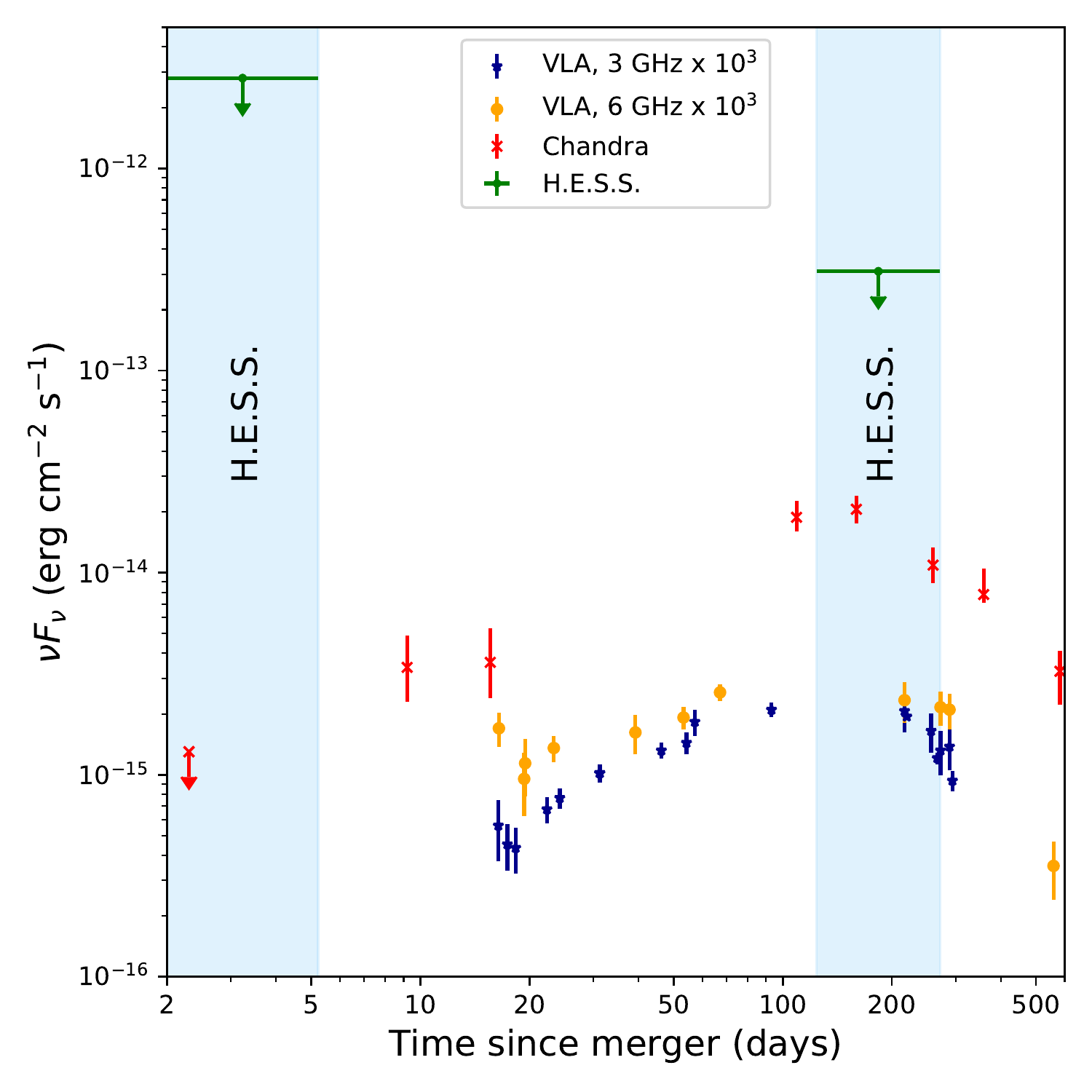}
\caption{Shown are the H.E.S.S. observation windows (blue areas), VLA
  radio data at 3 GHz (blue stars) and 6 GHz (orange circles)
  \citep{Hallinan2017, Alexander2017, Margutti2018, Dobie2018,
    Alexander2018, Mooley2018, Hajela:2019}, as well as X-ray data
  (red crosses) \citep[][and references therein]{Nynka2018,
    Hajela:2019}. The H.E.S.S. 1$-$10 TeV energy flux upper limits
  (green arrows) are derived for the prompt and the long-term
  follow-up with CT1-5 (c.f. Table~\ref{tab:data}).}
\label{fig:obs}
\end{figure}

After $\sim$160 days, the synchrotron radiation started to plateau and
later fade (Fig.~\ref{fig:obs}). This is similar to the behavior of a
young supernova remnant, and suggests that the ejecta is transitioning
from the free-expansion to the Sedov-Taylor phase when the ejected
mass of the merger remnant equals the swept-up circumstellar material.

\citet{Rodrigues2019} infer a total kinetic energy in the ejecta of
$10^{51}$\,erg, implying a total number of electrons in the remnant of
$10^{55}$. Accelerated electrons should also scatter off the intense
radio and X-ray synchrotron radiation field and produce synchrotron
self-Compton (SSC) emission. The expected peak energy of the SSC
component depends on the maximum accelerated electron energy as probed
by the X rays. While the radio to X-ray emission probes the product of
energy in electrons, $u_e$, and energy in magnetic fields, $u_B$, the
SSC component is proportional to $u_{e}^{2}\cdot u_B$. As shown by
\citet{Takami2014} and \citet{Rodrigues2019}, observations in the
$\gamma$-ray regime can break the ambiguity between $u_B$ and $u_e$
and provide crucial insight into the magnetic field in the ejecta as
well as the maximum accelerated particle energy. 

In this work, we present deep H.E.S.S. observations of GW170817 /
GRB~170817A covering the peak and onset of fading in the X-ray and
radio light-curves from 124 days to 272 days after the merger. This
measurement is accompanied by an improved analysis of the
H.E.S.S. data taken on the early (up to 5 days) kilonova. In the next
section we present the H.E.S.S. data set and results, followed by a
discussion on the implied magnetic field strength in a
non-relativistic kilonova scenario and a relativistic jet
scenario. Throughout this work we adopt a distance to the host galaxy
NGC\,4993 of 41.0\,Mpc, corresponding to a redshift of
$z=0.009727$~\citep{Hjorth2017a}\footnote{At this distance,
  very-high-energy (VHE; 100\,GeV$<$E$<$100\,TeV) photons pair-produce
  $e^{\pm}$ in interactions with the Extragalactic Background Light
  (EBL) on their way from GW170817 to Earth. The VHE flux reduction
  due to the EBL is energy-dependent and varies between 10\% and 30\%
  between 1\,TeV and 10\,TeV, respectively, assuming the
  \citet{Franceschini2008a} EBL model. Note that the model curves have
  been derived ignoring the EBL correction}.

\section{\label{sec:analysis}Data analysis and results}

\begin{table*}
\caption{\label{tab:data} Properties of the H.E.S.S. data sets on
  GW170817 / GRB~170817A and analysis results.}
\begin{ruledtabular}
\begin{tabular}{clccccccc}
 Data Set & Configuration & $T - T_0$ & Exposure & Energy range & $F(>E_\mathrm{th})$ & $F(1 - 10 \,\mathrm{TeV})$ & Zenith angle & Reference \\
 & & (days) & (hours) & TeV & erg\,cm$^{-2}$\,s$^{-1}$ & erg\,cm$^{-2}$\,s$^{-1}$ & $^{\circ}$& \\\hline
\RNum{1} & CT 5 & $0.22 - 5.23$ & 3.2 & 0.27$-$8.55 & $<1.5\times10^{-12}$ & & 58 & \citet{Abdalla2017a}\\
\RNum{2} & CT 1$-$5 & $0.22 - 5.23$ & 3.2 & 0.56$-$17.8 & $<4.7\times10^{-12}$ & $<2.8\times10^{-12}$ & 58 & this work\\
\RNum{3} & CT 1$-$5 & $124 - 272$ & 53.9 & 0.13$-$23.7& $<1.6\times10^{-12}$ & $<3.2\times10^{-13}$ & 24 & this work\\
\end{tabular}
\end{ruledtabular}
\end{table*}

The data set was obtained from observations with the H.E.S.S. phase II
array, including the upgraded 12\,m-diameter CT1-4 telescopes
\citep{HESS1U:2020} and the large 28\,m-diameter CT5 telescope. The
analysis presented by \citet{Abdalla2017a} used monoscopic data of the
28\,m telescope obtained between 5.3\,h and 5.3 days after the BNS
merger. Here we extend this analysis to also include data taken with
CT1-4. Observations from December 2017 to May 2018 with telescopes
pointing 0.5$^{\circ}$ offset from GW170817 were conducted allowing
for simultaneous estimation of the background level in the signal
region as discussed below. The different data sets are summarized in
Table~\ref{tab:data}. A standard data quality selection is applied to
the data \citep{Aharonian2006a, Abdallah2017a}. The events have been
selected and their direction and energy reconstructed using a
Monte-Carlo, template-based, shower model
technique~\citep{Parsons2014a}, requiring at least two telescope to
see the same $\gamma$-ray event. With this method, an energy
resolution of $\sim$10\% and angular resolution (at 68\% containment
radius) of 0.08$^{\circ}$ above $\gamma$-ray energies of 200\,GeV is
achieved. The resulting energy threshold of data set \RNum{3} is
$E_{\mathrm{th}} = 130$\,GeV. 
We define a circular region-of-interest centered on the optical
position of GW170817~\citep{Coulter2017a} with a radius of
0.09$^{\circ}$ for data sets \RNum{2} and \RNum{3} -- hereafter
referred to as the {\it ON region}. 10 to 20 background control
regions ({\it OFF regions}) are defined radially symmetric with
respect to the telescope pointing position for each
observation~\citep{Fomin1994}. This technique assures that the
$\gamma$-ray signal and background are estimated with the same
acceptance and under the same observation conditions.  
No significant $\gamma$-ray excess above the expected background is
detected from the direction of GW170817 in any data set. A second
analysis using an independent event calibration and reconstruction
\citep{de-Naurois2009a} confirms the result. A search for significant
emission on monthly timescales also does not reveal any signal. For
the total data set, 95\% confidence level (C.~L.) upper limits on the
photon flux are derived using the method described by
\citet{Rolke2005a} and assuming an underlying power-law spectral index
of the $\gamma$-ray emission of $\alpha=-2.0$
\citep[c.f.][]{Rodrigues2019}. Note that systematic errors are
subdominant compared to statistical uncertainties when deriving upper
limits, and are hence not considered here. All results are summarized
in Table~\ref{tab:data}. 

\begin{figure}[!h]
\centering
\includegraphics[width=0.483\textwidth]{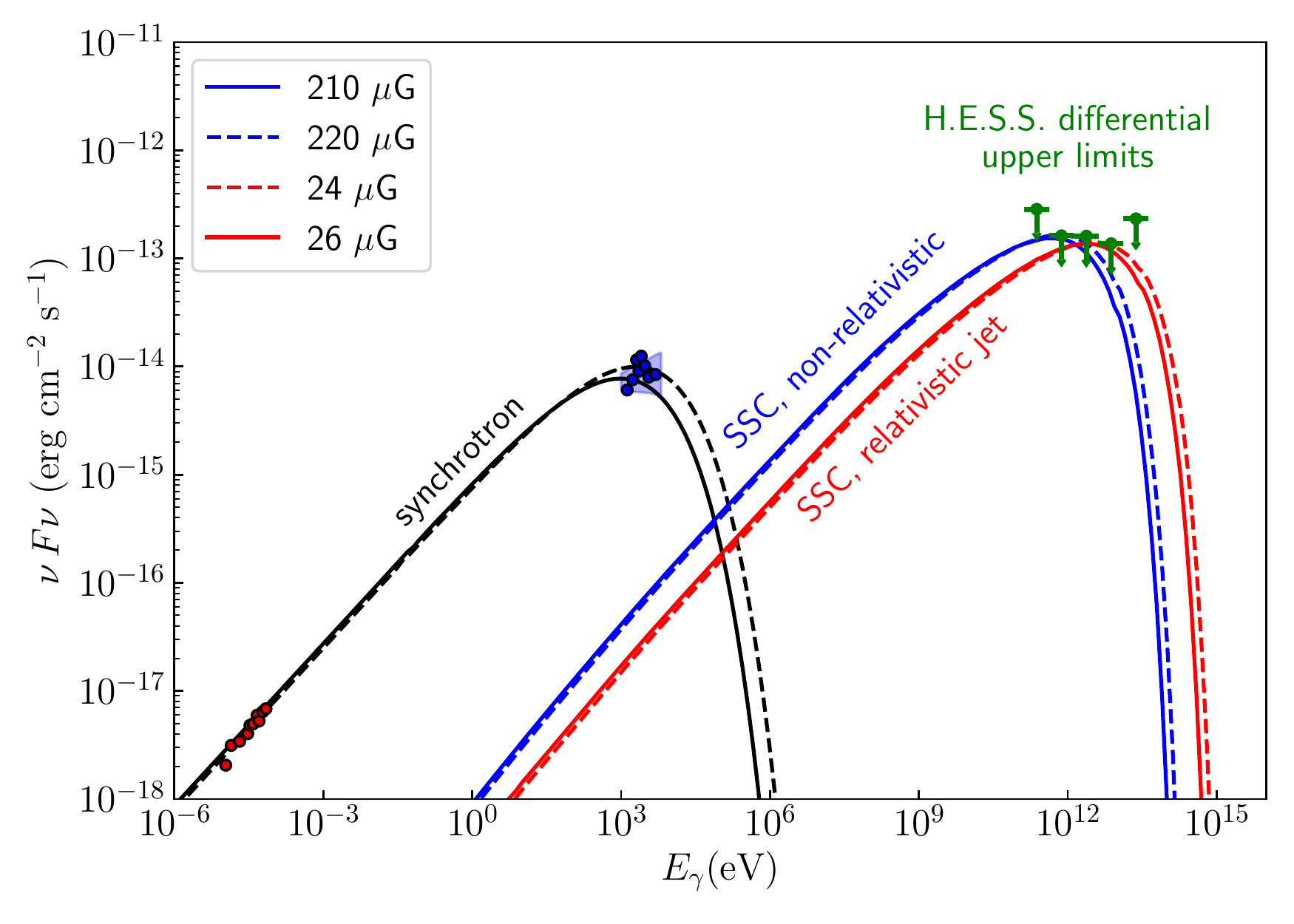}
\caption{Spectra predicted by the SSC modeling of the remnant of
  GRB~170817A, 110 days after the merger, for two distinct assumptions
  on the geometry and expansion speed of the remnant: an isotropic,
  non-relativistic expansion (blue SSC curves) and a relativistic jet
  (red SSC curves). For both assumptions, we show the minimum magnetic
  field strength imposed by the H.E.S.S. upper limits (green
  arrows). Solid and dashed curves are obtained by considering
  respectively the minimum and maximum flux values allowed by the
  X-ray measurements (blue points), while retaining compatibility with
  the radio data (red points).}
\label{fig:bmodel}
\end{figure}

Figure~\ref{fig:obs} depicts the radio and X-ray flux measurements
along with the inferred H.E.S.S. energy flux upper limits in the
$1-10$\,TeV energy range for data sets \RNum{2} and
\RNum{3}. Figure~\ref{fig:bmodel} shows the inferred energy flux upper
limits at the 95\% confidence level (C.~L.) in the VHE $\gamma$-ray
range from GRB~170817A for data set \RNum{3}, which has the best
sensitivity at TeV energies. In the next section we discuss how the
H.E.S.S. results constrain the magnetic field in the GW170817 ejecta
in a jet- and in a kilonova scenario. 

\section{\label{sec:discussion}Discussion}

Recent detections of VHE emission from GRBs over minutes
\citep{MAGIC:2019} and hours \citep{HESS:2019} motivate the search for
very late time emission from GRB-related events, like the remnant of
GW170817.
The H.E.S.S. differential upper limits can be translated into an
integral energy flux limit, for a given assumption on the spectrum of
the radiating particles. In turn, this limit provides a constraint on
the magnetic field strength under the assumption of one-zone
synchrotron emission with corresponding inverse Compton (IC) emission.

Observational evidence suggests that at early times, the kilonova provides the dominant target
radiation field in the remnant \citep{Villar:2017wcc}. However, the
decay of this component, whose flux falls steeply with $t^{-2.3}$,
results in a late-time dominance of the synchrotron radiation in the
source. It is therefore naturally expected that at late times SSC will
dominate the remnant's IC emission.

The measured X-ray flux of the source can be used to infer the X-ray
luminosity emitted as synchrotron radiation, $L_X$. In order to
consistently model the emission from this electron population, a
geometry assumption is necessary. We consider two scenarios: one where
the remnant expands isotropically and non-relativistically, and the
other where a relativistic jet is launched.

In the isotropic scenario, we assume a volume-filled spherical emitter
with radius $R_\mathrm{iso}=\beta c\Delta t$, where $\Delta
t=110~\mathrm{days}$ is the time since the merger 
\footnote{We base this discussion on the 110 day timescale, for which
  there are quasi-simultaneous flux measurements in both radio and X
  ray bands. As shown in Fig.~\ref{fig:obs}, the flux levels at this
  timescale are comparable throughout the H.E.S.S. observation
  window.}. Based on photospheric velocity measurements of the remnant
\citep{Piro:2017ayh}, we consider a value of the expansion speed of
$\beta=0.2$. 

In the relativistic scenario, we consider a jet with speed
$\beta=0.94$ (corresponding to a Lorentz factor of $\Gamma=3$) at late
times, and a jet opening angle\footnote{Throughout this text, primed
  quantities denote parameter values in the shock rest frame and
  unprimed ones in the observer's frame.} of
$\theta^{'}_{\mathrm{jet}}=5^\circ$, observed at an angle of
$\theta_\mathrm{obs}=20^\circ\approx180^\circ/\pi\Gamma$ from the jet
axis; the source is therefore observed with a doppler factor
$\delta=\Gamma=3$. These parameter values are motivated by radio
observations of superluminal motion of the source
\citep{Mooley:2018dlz}. Furthermore, from a purely theoretical
standpoint, this value of the Lorentz factor is expected at $\sim100$
day timescales, considering the energy transfer from the shock into a
surrounding medium of constant density~\citep{Rees:1992ek}. 
The emitting region is assumed to be a spherical blob at the front
edge of the jet, whose radius is therefore given by
$R^\prime_\mathrm{blob} \approx \delta \beta c \Delta t
\theta^\prime_\mathrm{jet}$. 

The maximum energy of the emitted synchrotron radiation is fixed by
X-ray observations, $E_{\mathrm X}\approx
10~\mathrm{keV}$~\citep{Nynka2018}. This is related to the magnetic
field strength in the source, $B^\prime$, and the maximum electron
energy of the emitting electrons, $E^\prime_e=\delta^{-1}E_e$, through 
$E^\prime_{\mathrm X} \propto E^{\prime2}_e B^\prime$.
This means that in the relativistic scenario, for a given value of
$B^\prime$, the maximum energy of the emitting electrons scales as  
$E^\prime_e\propto\delta^{-0.5}$.
Furthermore, since the jet blob emits isotropically in its own rest
frame, the X-ray luminosity, obtained assuming the observed emission
is isotropic, $L^{\mathrm{iso}}_X$ deduced from flux measurements
relates to the luminosity in the rest frame of the blob $L_X^\prime$
through $L^{\mathrm{iso}}_\mathrm{X}=\delta^{4}L^\prime_{X}$. This
luminosity relates to the total number of X-ray emitting electrons,
$N^\prime_{e}$, as well as $E_{e}^\prime$ and  $B^\prime$, through 
$L^{\prime}_\mathrm{X}=N_{e}^\prime
E_{e}^\prime/\tau^\prime_{\mathrm{syn}}(E^\prime_{e}) \propto
N_{e}^\prime E_{e}^{\prime2}B^{\prime2}$, 
where $\tau^\prime_\mathrm{syn}(E_e^\prime)$ is the synchrotron
cooling timescale. 
Therefore, in the relativistic jet scenario, for a given value of
$B^\prime$ and measured X-ray flux, the number of electrons scales as
$N^\prime_e\propto\delta^{-3}$. This will affect the results of the
SSC model, since for higher values of $\delta$ there will be fewer
emitting electrons, and a lower density of synchrotron photons, thus
reducing the expected $\gamma$-ray flux. 

The $\gamma$-ray luminosity expected from SSC is given in the shock
rest frame by $L^\prime_{\mathrm{IC}}= N_{e}^\prime
E^\prime_{e}/\tau^\prime_{\mathrm{IC}}(E^\prime_{e})\propto
N_{e}^{\prime}(L^\prime_{\mathrm{X}}/(4\pi R^{\prime2}
c))E_{e}^{\prime0.5}$, where $\tau^\prime_{\mathrm{IC}}(E^\prime_{e})$
is the IC cooling timescale, and $R^{\prime}$ is the size of the
emitting region (either $R^\prime_\mathrm{blob}$ or
$R^\prime_\mathrm{iso}$, defined above). 
The approximate scaling with $E_{e}^{\prime0.5}$ is due to the fact
that the IC cooling occurs deep in the Klein-Nishina regime (since the
observed synchrotron spectrum peaks at X-ray energies and the Lorentz
factor of the ejecta is mild). Putting together the scalings given
previously, we obtain that for a given value of $B^\prime$, the IC
emission for the relativistic jet scenario scales as
$L^\prime_{IC}\propto \theta_\mathrm{jet}^{\prime-2}(\beta\Delta
t)^{-2}\delta^{-9.25}$ in the shock rest frame, which corresponds to a
luminosity in the observer's frame of 

\begin{equation}
    L_{IC}\approx10^{46}\left(\frac{\theta^\prime_\mathrm{jet}}{5^\circ}\right)^{-2}\left(\frac{\beta\Delta t}{0.94\times110~\mathrm{days}}\right)^{-2}\left(\frac{\delta}{3}\right)^{-5.25}~\mathrm{erg/s}.
    \label{equ:scaling}
\end{equation}

Thus, a slower expansion of the emission region would lead to a more
compact source and therefore a stronger constraint on the magnetic
field strength (both in the jet-like and isotropic scenarios). 

In Fig.~\ref{fig:bmodel} we show the modeled synchrotron emission
spectrum (black curves) and the respective SSC emission for both
scenarios introduced above. These spectra were obtained with a
numerical radiation model, introduced by \citet{Rodrigues2019}. In
this model, a population of electrons is considered to fill
homogeneously the emission region, and to be continuously accelerated
to  a power-law spectrum, with a possible cut-off at the highest
energies. As can be seen in Fig.~\ref{fig:bmodel}, these
characteristics can explain radio (red points) and X-ray observations
(blue points and blue shaded region). The parameters of the electron
population have then been adjusted, and the magnetic field strength
minimized, so that the predicted SSC emission does not exceed any of
the 95\% C.L. upper limits in the VHE $\gamma$-ray range derived from
the H.E.S.S. observations (green). 
The solid vs. dashed curves represent the two extreme cases consistent
with the observed X-ray and radio fluxes. In the non-relativistic
scenario, the H.E.S.S. limits can constrain the minimum magnetic field
strength to  $\gtrsim210\,\mu\mathrm{G}$. In contrast, in more highly
relativistic scenarios, the lower limit is weakened to the level of
$\gtrsim24\,\mu\mathrm{G}$. The reason is that the strong scaling of
the SSC flux with the Doppler factor (Eq.~\ref{equ:scaling}) implies
that for a more relativistic outflow, a higher electron number is
necessary to reach the flux limits, which implies a lower magnetic
field in order to maintain the X-ray flux.  

As a point of comparison with the lower limit obtained through this
analysis, the minimum magnetic field expected at late times downstream
of the shock is of the order of
$\sim100(\Gamma/3)(B_\mathrm{ISM}/10~\mu\mathrm{G})~\mu\mathrm{G}$
\citep{Kumar:2012xm}, where $B_\mathrm{ISM}$ is the magnetic field
strength in the interstellar medium (ISM). Furthermore, observations
of the prompt emission of GRB 080916C by Fermi-LAT and 1-day afterglow
emission in X rays and optical wavelengths have provided evidence of
magnetic fields in the shock that are at the level of the compressed
surrounding medium \citep{Kumar:2009ps}, thus suggesting an ISM
magnetic field of $\sim10~\mu\mathrm{G}$. 

Very high-energy $\gamma$-ray observations can provide a direct probe
of the magnetic field in BNS merger remnants. While the radio and
X-ray data constrain the synchrotron part of the non-thermal spectrum,
a measurement of the IC component in $\gamma$ rays is needed to break
the ambiguity between energy in electrons and magnetic
fields. Interferometric radio observations and long-term X-ray
observations are crucial for inferring the jet properties such as the
opening angle, viewing angle, or the Doppler factor. Based on the
GW~170817 / GRB~170817A characteristics, one other short GRB may have
been seen off-axis (GRB~150101B; \citep{Troja2018}). 
Long-term monitoring of BNS mergers in the radio, X-ray and
$\gamma$-ray domain is necessary to further constrain the source
properties and bridge the gap between the early-time kilonova and
non-thermal GRB emission on the one hand, and the long-term behavior
and the interaction between the jet and the ISM on the other hand.
Current-generation Imaging Atmospheric Cherenkov Telescopes (IACTs)
such as H.E.S.S., MAGIC or VERITAS can search for and study
$\gamma$-ray counterparts on days-to-month timescales for BNS merger
remnants seen under different viewing angles.
Furthermore, the future Cherenkov Telescope Array (CTA) will be an
order of magnitude more sensitive at around 1\,TeV, allowing us to
constrain the minimum magnetic field in events like GW170817 /
GRB~170817A to the mG regime. This will allow CTA to detect VHE
$\gamma$-ray emission from events such as GW170817.
As noted previously by \cite{Rodrigues2019}, future observations may
also be able to better constrain the nature of the non-thermal
electrons. One possibility is that the observed radiation is dominated
by ``fresh'' electrons picked up from the surroundings of the merger
and accelerated at the shock front, allowing for hard photon spectra
into the gamma-ray range. In a different scenario, the emission might
originate in ``old'' electrons, continuously accelerated inside the
volume of the ejecta, leading to cooling features at high energies. 

\section{Acknowledgments}
The support of the Namibian authorities and of the University of
Namibia in facilitating the construction and operation of H.E.S.S. is
gratefully acknowledged, as is the support by the German Ministry for
Education and Research (BMBF), the Max Planck Society, the German
Research Foundation (DFG), the Helmholtz Association, the Alexander
von Humboldt Foundation, the French Ministry of Higher Education,
Research and Innovation, the Centre National de la Recherche
Scientifique (CNRS/IN2P3 and CNRS/INSU), the Commissariat \`a
l'\'Energie atomique et aux \'Energies alternatives (CEA), the
U.K. Science and Technology Facilities Council (STFC), the Knut and
Alice Wallenberg Foundation, the National Science Centre, Poland grant
no. 2016/22/M/ST9/00382, the South African Department of Science and
Technology and National Research Foundation, the University of
Namibia, the National Commission on Research, Science \& Technology of
Namibia (NCRST), the Austrian Federal Ministry of Education, Science
and Research and the Austrian Science Fund (FWF), the Australian
Research Council (ARC), the Japan Society for the Promotion of Science
and by the University of Amsterdam. We appreciate the excellent work
of the technical support staff in Berlin, Zeuthen, Heidelberg,
Palaiseau, Paris, Saclay, T\"ubingen and in Namibia in the construction
and operation of the equipment. This work benefitted from services
provided by the H.E.S.S. Virtual Organisation, supported by the
national resource providers of the EGI Federation. XR was supported by
the Initiative and Networking Fund of the Helmholtz Association. 

\bibliography{EM170817_HESS}
\end{document}